# Binary Image-Based Intrusion Detection for Operational Technology Networks: Extending the SPHBI Methodology from IoT to Modbus TCP

Aamir Omar[0009-0003-7802-2277]

[1] BMT, London, UK
aamir.omar@uk.bmt.org

**Abstract.** This paper extends the Single Packet Header Binary Image (SPHBI) intrusion detection methodology from IoT to Modbus TCP, evaluating five approaches spanning a gradient of protocol depth on the CIC Modbus 2023 dataset (11.4 million packets, eight detectable attack types). TCP/IP headers alone achieve only 51.8% binary accuracy, confirming that header-level heterogeneity exploited in IoT traffic is absent in uniform SCADA environments. Adding eight bytes of application-layer information improves binary accuracy to 98.1% with just 63 parameters, directly relevant to per-packet classification on resource-constrained OT edge devices. The best-performing approach achieves 94.4% ± 2.2pp multiclass accuracy across nine classes (95% CI [92.9%, 95.9%], 10 seeds) with 56,873 parameters, roughly 430 times fewer than comparable ResNet50-based approaches. Per-class recall analysis shows seven of eight detectable attack types identified with recall above 94%, while replay attacks remain structurally undetectable by any single-packet method.

## 1 Introduction

Operational technology (OT) networks underpin critical infrastructure across energy, water, manufacturing and transport sectors. These networks rely on industrial control systems (ICS) and supervisory control and data acquisition (SCADA) architectures to monitor and control physical processes, typically using legacy protocols such as Modbus TCP and DNP3. Historically isolated from corporate IT networks, OT environments have become increasingly interconnected as organisations pursue remote monitoring and cloud integration (Termanini et al., 2024). This convergence has expanded the attack surface considerably. Incidents including the 2015 Ukrainian power grid attack, the Stuxnet worm and the 2017 TRITON malware campaign demonstrated that adversaries can exploit OT protocol vulnerabilities to cause physical damage and disrupt essential services (Kotsiopoulos et al., 2025). Automated per-packet intrusion detection reduces dependence on human vigilance by providing continuous monitoring that operates independently of operator awareness.

Most existing machine learning-based IDS for OT rely on extracted flow-level features or handcrafted protocol fields, requiring domain-specific feature engineering that limits portability across protocols (Termanini et al., 2024; Anthi et al., 2021). Flow-based approaches aggregate multiple packets into statistical summaries, introducing latency and sacrificing per-packet granularity needed for real-time detection in safety-critical environments.

El-Sherif, Khattab and El-Soudani (2025) proposed the Single Packet Header Binary Image (SPHBI) methodology, which extracts 18 bytes from TCP/IP packet headers, converts them into a 12×12 binary image and classifies the image using lightweight CNNs. The binary classifier totals 35 trainable parameters and achieves 100% binary accuracy on the Edge-IIoTset dataset (Ferrag et al., 2022). By contrast, Kotsiopoulos et al. (2025) applied a ResNet50-based CNN to Modbus TCP with over 23 million parameters, roughly 660,000 times larger. Such models are impractical for deployment on the embedded devices and PLCs found in OT environments. A binary classifier small enough for edge deployment would enable per-packet screening at the network boundary between SCADA master and field devices, flagging malicious traffic before it reaches the target IED.

Despite these results, SPHBI has not been evaluated on OT protocol traffic. This gap is significant because IoT traffic exhibits considerable header-level diversity (varied packet sizes, ports, TTL values), whereas Modbus SCADA traffic is highly homogeneous: a single master polls a small number of IEDs over port 502 with near-identical header configurations. The CIC Modbus 2023 dataset (Boakye-Boateng, Ghorbani and Lashkari, 2023) provides no packet-level labels, and prior classification results on it are limited to Russo, Zanasi and Marasco (2024), who achieved 84.2% binary accuracy using Zeek-extracted flow features.

This paper addresses these gaps through a systematic comparison of five approaches representing a gradient of protocol depth, from TCP/IP headers only to application-layer fields with PDU operands. Table 1 summarises the approaches and headline results.

**Table 1.** Summary of five experimental approaches. Multiclass accuracy is mean ± std across 10 seeds. Binary accuracy is mean ± std across 10 seeds.

| App. | Protocol layers | Bytes | Image | Bin. acc. | MC acc. | Params (binary) |
|---|---|---|---|---|---|---|
| 1 | TCP/IP only | 18 | 12×12 | 51.8±0.0% | 45.5±5.7% | 51 |
| 2 | TCP/IP+MBAP+FC | 26 | 16×13 | 98.1±0.3% | 87.6±4.9% | 63 |
| 2b | TCP/IP+MBAP+FC+PDU | 30 | 16×15 | 98.6±0.0% | 94.4±2.2% | 73 |
| 3 | App. layer only | 8 | 8×8 | 93.7±4.8% | 84.6±8.5% | 38 |
| 3b | App. layer+PDU | 12 | 12×8 | 98.6±0.0% | 94.5±4.0% | 39 |

The principal contributions are: (1) the first application of single-packet binary image-based deep learning to Modbus TCP traffic, achieving 98.6% binary accuracy with 73 parameters and 94.4% ± 2.2pp multiclass accuracy with 56,873 parameters; (2) a systematic five-approach comparison quantifying the discriminative contribution of each protocol layer; (3) a transparent, reproducible labelling methodology for the CIC Modbus 2023 dataset; and (4) a detailed analysis of attack detectability limits.

The single-packet paradigm has inherent detection boundaries. Replay attacks retransmit copies of legitimate packets with identical headers, and delay response attacks modify only timing. Neither is detectable by any method that examines individual packets in isolation. These limits are analysed in Section 6 and inform the scope of claims throughout this paper.

# 2 Related Work

## 2.1 Visual and Binary Image-based Classification

The idea of representing network traffic as images for classification has been explored through several approaches. Nataraj et al. (2011) classified malware binaries as greyscale images (98.0% accuracy), Wang et al. (2017) applied CNNs to greyscale packet representations (89.9% on UNSW-NB15), and Baptista, Shiaeles and Kolokotronis (2019) used Hilbert curve colour images with SOINN (94.1%). These operated on aggregated flows or files rather than individual packets and were evaluated on IT or malware datasets, not OT traffic.

The SPHBI methodology (El-Sherif, Khattab and El-Soudani, 2025), described in Section 1, achieves its strong results on IoT traffic precisely because heterogeneous devices produce rich header-level variation. OT traffic lacks this variation. The question is not whether CNNs can classify binary images (this is established) but whether binary images derived from uniform OT headers contain sufficient discriminative information to support classification, and if not, which additional protocol layers restore it. This paper provides the first empirical answer for Modbus TCP.

Kotsiopoulos et al. (2025) applied visual binary representations to Modbus TCP, the only prior use of image-based classification in OT. Their system converts bidirectional flows into colour images via Hilbert curves and classifies them using a ResNet50-based CNN (98.4% accuracy, 23M+ parameters). This approach differs from SPHBI in operating on aggregated flows (precluding per-packet guarantees), using a model three orders of magnitude larger, and evaluating on a custom dataset rather than a public benchmark.

## 2.2 ICS Intrusion Detection and the CIC Modbus 2023 Dataset

The dominant ICS IDS paradigm extracts features using tools such as Zeek or CICFlowMeter and applies classifiers including random forests and deep learning architectures (Termanini et al., 2024). This creates two problems: protocol-specific expertise is required, limiting portability, and flow aggregation introduces detection latency. Anthi et al. (2021) proposed a three-tiered ICS IDS and found that function code, CRC and Modbus frame length are the most important features for Modbus attack detection. This independently confirms that application-layer fields carry the primary discriminative signal, consistent with our Approach 1 failure (51.8%).

The CIC Modbus 2023 dataset (Boakye-Boateng, Ghorbani and Lashkari, 2023) simulates a Docker-based electrical substation with Modbus TCP communication. It includes nine attack types across three scenarios but provides no packet-level labels. The dataset reproduces characteristics representative of production SCADA: continuous polling during attacks, extreme class imbalance and realistic function code usage. It does not reproduce concurrent non-Modbus protocols, timing jitter or environmental noise. Russo, Zanasi and Marasco (2024) achieved 84.2% binary accuracy using Zeek flow features and an LSTM. No prior work has applied single-packet binary image classification to this dataset.

# 3 Methodology

## 3.1 Dataset and Extraction

Packet extraction used tshark 4.4.14 in a Docker container, extracting 26 fields per Modbus TCP packet. Capture point selection was conducted empirically by surveying all 24 PCAP directories. The selected capture points yielded 4,101,373 benign packets (19 files), 282,078 external attack packets (2 files), 3,657,304 compromised SCADA packets (17 files) and 3,348,642 compromised IED packets (20 files), totalling 11,389,397 Modbus packets across 58 files. A systematic timestamp offset analysis confirmed zero offset between PCAPs and attack logs for all scenarios.

## 3.2 Labelling Methodology

Since 94% of traffic within attack captures is benign polling, a labelling methodology is required to identify malicious packets. Three strategies were evaluated. Pure timestamp windowing was rejected (it would mislabel benign polling as attacks). Generic feature filtering was rejected as insufficiently specific. Strategy 3 (hybrid per-attack-type rules) was chosen, combining attack log time windows with protocol-level packet signatures. Table 2 summarises the rules.

**Table 2.** Per-attack-type labelling rules and resulting packet counts. Total: 10,747,066 Normal, 642,331 Attack.

| Attack type | Labelling rule | Packets |
|---|---|---|
| Brute force | Within window AND func_code = 5 | 632,450 |
| Frame stacking | Self-identifying: comma-separated function codes | 1,335 |
| Query flooding | All Modbus packets within window | 5,967 |
| Reconnaissance | All Modbus packets within window | 355 |
| Replay | All Modbus packets within window | 1,782 |
| Payload injection | All packets within window | 363 |
| Length manip. | Within window AND anomalous (func_code, mbtcp_len) | 26 |
| FDI | Self-identifying: byte_cnt = 171 | 53 |
| Delay response | Excluded (undetectable) | — |

Domain knowledge was used only for labelling, not for feature engineering. The CNN receives raw binary images and learns its own discriminative features. The rules are documented so their effect on ground truth can be assessed independently. Edge cases including missing completion markers in the IED scenario, unclosed attack windows and overlapping attack types were handled through documented heuristics.

## 3.3 Byte Reconstruction and Approach Design

The application tshark outputs decoded field values rather than raw bytes (e.g. ip.hdr_len=20 instead of IHL nibble=5). A reconstruction script reverses each decoding using bitwise operations to recover the original byte values. Multi-byte fields are split into high and low bytes in network byte order. The reconstruction was validated by confirming all 11.4 million packets produced values in [0, 255], version/IHL = 0x45 for 100% of packets and protocol = 6 (TCP) throughout.

The 30 reconstructed bytes are arranged into five approaches representing a gradient of protocol depth (Table 1). Each approach extracts a contiguous subset and converts it into a binary image by unpacking each byte into 8 bits. Figure 1 illustrates the byte layout.

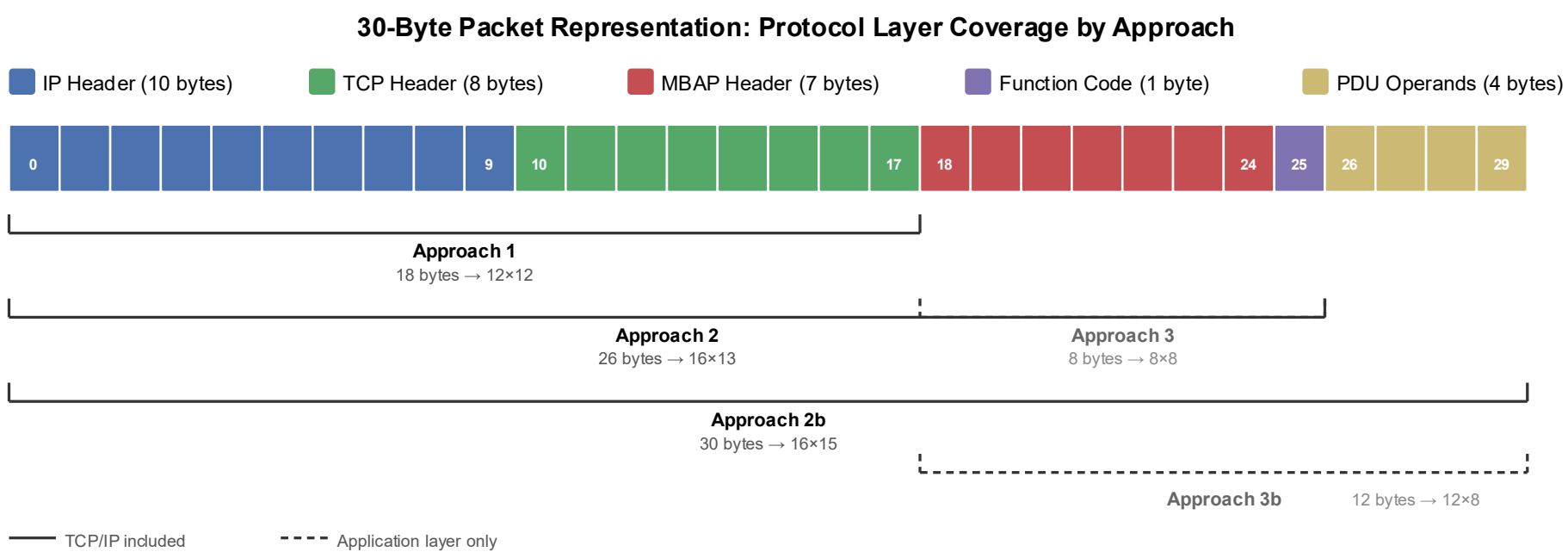

**Fig. 1.** Byte layout and approach coverage. Each of the 30 byte positions is colour-coded by protocol layer. Solid brackets denote approaches that include TCP/IP headers; dashed brackets denote application-layer-only approaches.

## 3.4 CNN Architecture

The binary classifier follows the SPHBI architecture: two convolutional layers with one filter each, Tanh activation and MaxPool(2, stride=2). Kernel sizes adapt dynamically to input dimensions. Parameter counts range from 38 (Approach 3) to 73 (Approach 2b).

The multiclass classifier uses two convolutional layers with 32 filters each (3×3 with padding=1, then 2×2), Sigmoid activation and MaxPool(2, stride=1). The flatten dimension is computed via a dummy forward pass, allowing the same architecture to handle all five image sizes. Class-weighted cross-entropy loss addresses class imbalance, with weights computed as N/(K×N_c). Training data is capped at a configurable maximum per class; the test set is never capped. Figure 2 illustrates the multiclass architecture for Approach 2b.

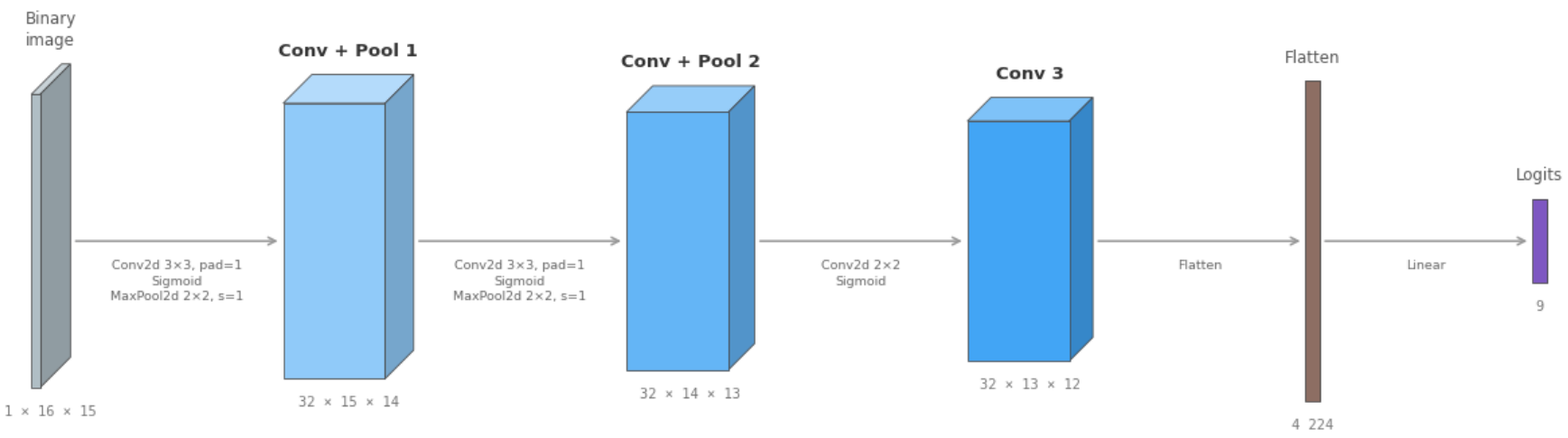

**Fig. 2.** Multiclass CNN architecture for Approach 2b (16×15 input, 56,873 parameters)

The original SPHBI multiclass classifier uses four convolutional layers with Sigmoid activation. When applied to input images larger than 8×8, this architecture suffered from vanishing gradients: training loss remained effectively constant across 100 epochs and the model collapsed to predicting the majority class. All approaches were standardised to the two-layer architecture, which resolved this and ensures that performance differences reflect the information content of the input rather than architectural artefacts.

Figure 3 shows representative binary images for each traffic class across all five approaches. The visual uniformity within Approach 1 rows illustrates why the model cannot discriminate classes using TCP/IP headers alone. By Approaches 3 and 3b, structural differences between normal polling and attack traffic become visually apparent.

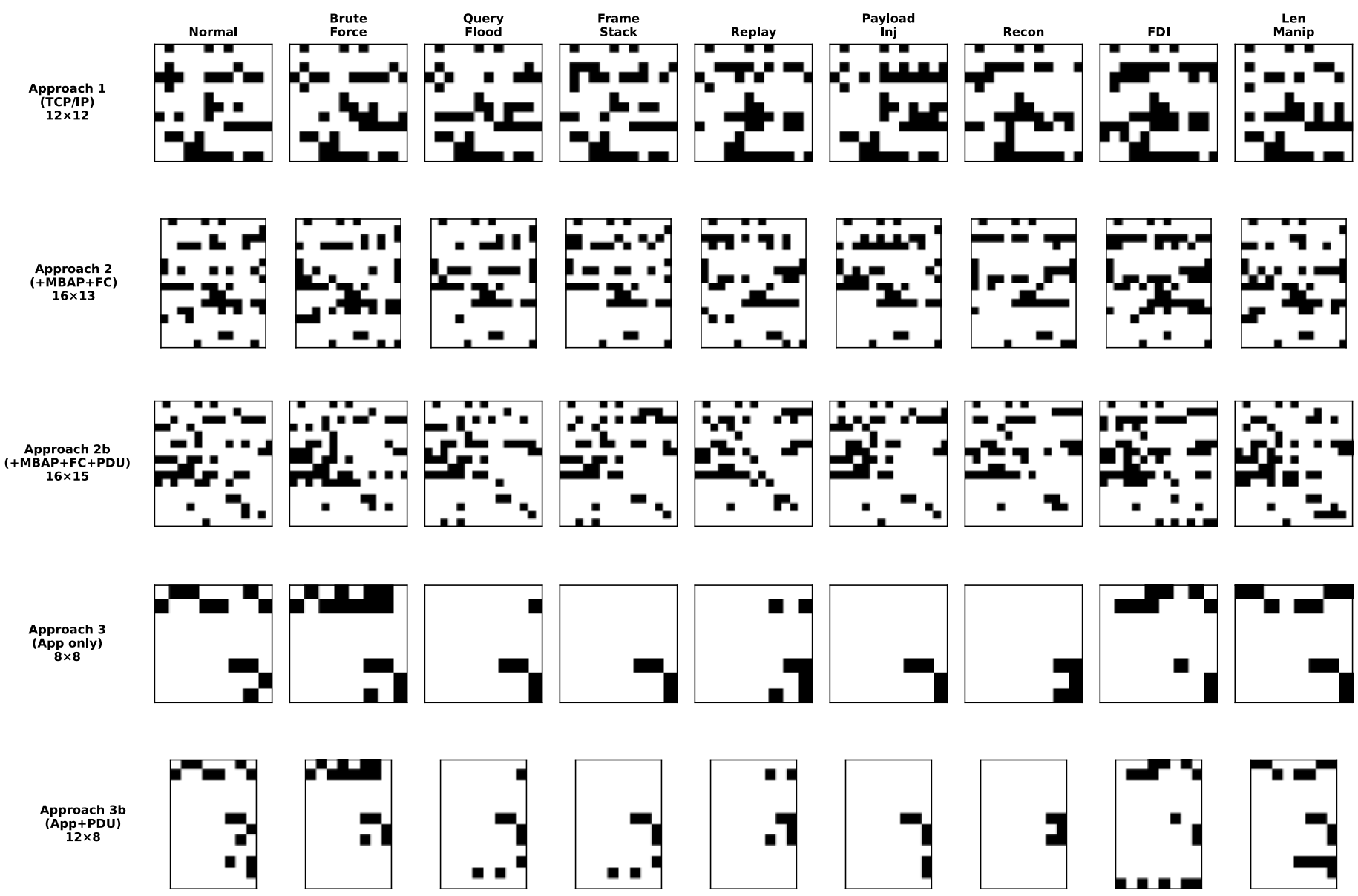

**Fig. 3.** Binary image representations of sample Modbus packets for nine traffic classes (columns) across five byte-selection approaches (rows). Each packet's raw bytes are unpacked into bits and reshaped into a 2-D grid, displayed with black pixels representing bit value 1 and white pixels representing bit value 0. The varying approaches illustrate how attack signatures manifest at different protocol layers.

## 4 Experimental Setup

The dataset was split 80/10/10 (train/validation/test) using stratified sampling by multiclass label by percent rank within each class. Table 3 shows the class distribution. Normal traffic constitutes 94.4% of the total. Six of nine classes have fewer than 600 test samples. FDI (6 test samples), length manipulation (3), recon (36) and payload injection (37) are included experimentally; per-class metrics for these classes should be interpreted as indicative rather than statistically robust.

**Table 3.** Class distribution across train/val/test sets (80/10/10 stratified).

| Class | Train | Val | Test | % total |
|---|---|---|---|---|
| Normal | 8,597,652 | 1,074,707 | 1,074,707 | 94.36 |
| Brute force | 505,960 | 63,245 | 63,245 | 5.55 |
| Query flooding | 4,773 | 597 | 597 | 0.052 |
| Replay | 1,425 | 178 | 179 | 0.016 |
| Frame stacking | 1,068 | 133 | 134 | 0.012 |
| Payload inj. | 290 | 36 | 37 | 0.003 |
| Recon | 284 | 35 | 36 | 0.003 |
| FDI | 42 | 5 | 6 | <0.001 |
| Length manip. | 20 | 3 | 3 | <0.001 |

Majority classes are capped at 5,000 (multiclass) or 50,000 (binary) samples per class. Inverse-frequency class weights are applied to the loss function: weight_c = N/(K×N_c). The validation set is capped identically and used for model selection; the test set is never capped.

Three experimental phases were conducted. Phase 1 (cross-approach comparison) trained all five approaches on both tasks with identical hyperparameters: multiclass used SGD (momentum 0.9, lr=0.01, 100 epochs, batch 32); binary used Adam (lr=0.01, 20 epochs, batch 32). Phase 2 (hyperparameter sensitivity) ran a $2^4$ factorial experiment on Approach 2b (32 runs). Phase 3 (cap sensitivity) swept seven cap values (500 to 50,000) for Approaches 2b, 3 and 3b. Multi-seed validation (10 seeds) was conducted for Phases 1 and 2 and the binary cross-approach comparison. Results are reported as mean ± std with 95% confidence intervals. Pairwise significance was assessed using Welch's t-test ($p < 0.05$). All experiments ran on a single NVIDIA T4 GPU.

# 5 Results

## 5.1 Five-approach Comparison

The binary results (Table 1) reveal a sharp divide. Approach 1 achieves only 51.8% accuracy with 10.5% attack precision. Adding eight bytes of application-layer information (Approach 2) restores accuracy to 98.1%. Approaches 2, 2b and 3b are stable across 10 seeds (std ≤ 0.3pp), confirming that binary classification is reliably addressed by the single-packet binary image approach when application-layer fields are included. Approach 3 is the exception: its mean is 93.7% ± 4.8pp (range 89.0–98.5%), exhibiting the same instability observed in the multiclass task.

Table 4 presents the multiclass results with confidence intervals. Approach 2b achieves the highest mean accuracy (94.4% ± 2.2pp) with the narrowest confidence interval. Approach 3b is statistically indistinguishable (94.5% ± 4.0pp, $p = 0.929$) but with nearly twice the variance. Approach 2 is significantly lower (87.6%, $p = 0.002$). Approach 3 (84.6% ± 8.5pp) shows the highest instability, with individual seeds ranging from 64.2% to 94.4%.

**Table 4.** Phase 1 multiclass results (10 seeds per approach, cap = 5,000). Approaches 2b and 3b are statistically indistinguishable ($p = 0.929$).

| App. | Protocol depth | Acc. (mean±std) | 95% CI | Params | Notes |
|---|---|---|---|---|---|
| 1 | TCP/IP only | 45.5±5.7% | [41.4, 49.6] | 33,257 | Near-random |
| 2 | TCP/IP+MBAP+FC | 87.6±4.9% | [84.1, 91.1] | 48,809 | — |
| 2b | TCP/IP+MBAP+FC +PDU | 94.4±2.2% | [92.9, 95.9] | 56,873 | Lowest variance |
| 3 | App. layer only | 84.6±8.5% | [78.5, 90.7] | 14,825 | High variance |
| 3b | App. layer+PDU | 94.5±4.0% | [91.6, 97.4] | 21,737 | ≈ 2b (p=0.929) |

Approach 1's sole success is frame stacking (97.8% recall), which modifies TCP-level packet length. This confirms the CNN extracts whatever signal is present; the failure reflects absent information, not model inadequacy. PDU operands add significant value: Approach 2b improves on Approach 2 by 6.8pp ($p = 0.002$), and Approach 3b outperforms Approach 3 by 9.9pp ($p = 0.006$). The statistical equivalence of Approaches 2b and 3b ($p = 0.929$) has a practical implication: the TCP/IP headers, which constitute 18 of Approach 2b's 30 input bytes, contribute no statistically detectable improvement over the 12-byte application-layer representation at this operating point. The discriminative information for Modbus intrusion detection resides almost entirely in the application layer.

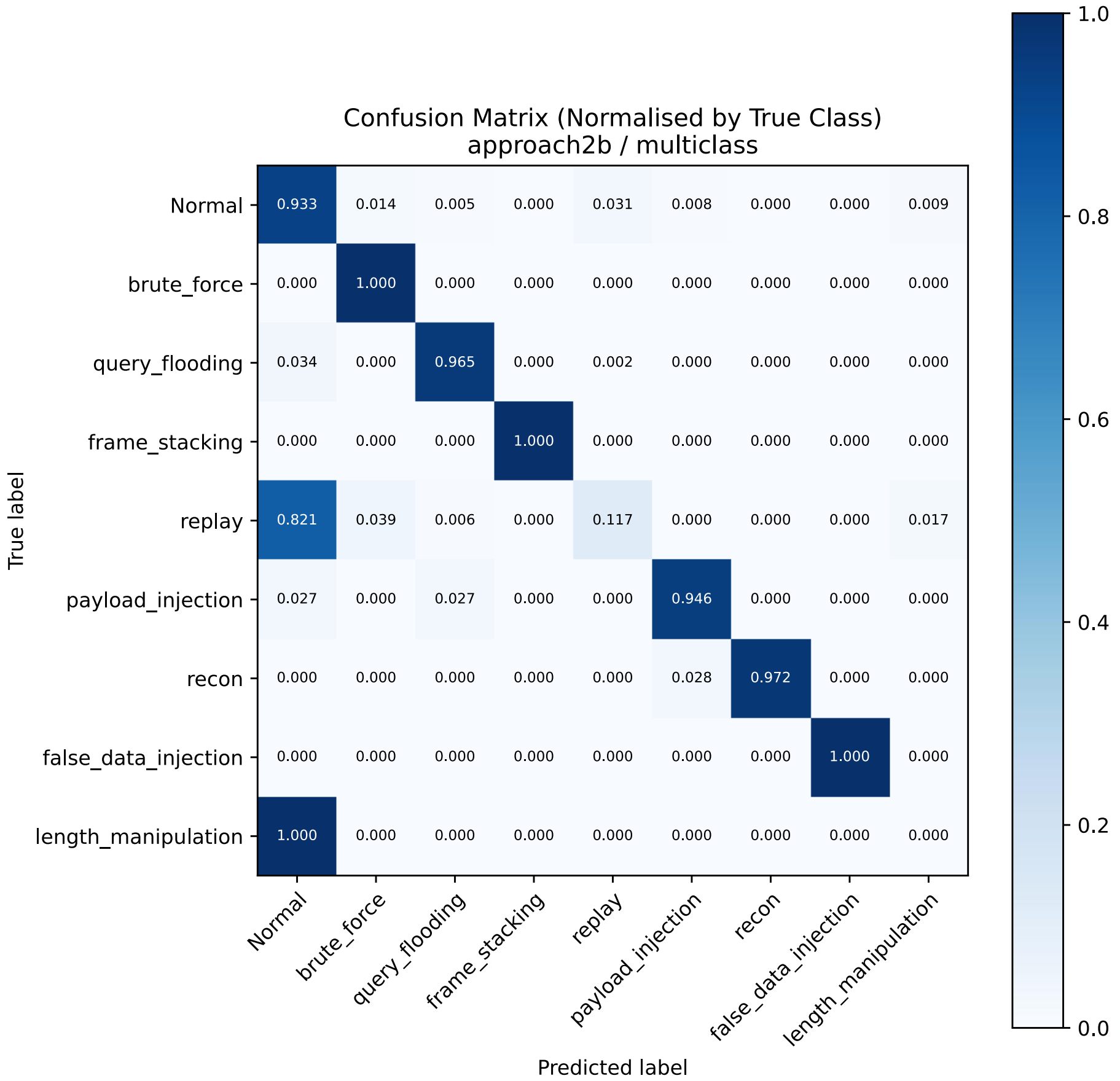

**Fig. 4.** Normalised confusion matrix for Approach 2b multiclass (cap = 5,000). Rows represent true labels; columns represent predicted labels. Replay and length manipulation are the only classes with recall below 50%.

### 5.2 Per-class Analysis

Figure 4 shows the normalised confusion matrix for Approach 2b at the recommended operating point (cap = 5,000). Across 10 seeds, seven of eight detectable attack types achieve mean recall above 94%: brute force (100.0% ± 0.0), frame stacking (99.0% ± 1.5), FDI (100.0% ± 0.0), reconnaissance (98.3% ± 1.4), query flooding (95.9% ± 0.7), payload injection (94.6% ± 0.0) and Normal (94.1% ± 2.3). Only replay (7.9% ± 4.5) and length manipulation (3.3% ± 10.5, not distinguishable from zero) fail. The near-zero standard deviation for payload injection and FDI indicates the CNN finds the same decision boundary regardless of initialisation, consistent with unambiguous attack signatures.

**Table 5.** Per-class recall for Approach 2b multiclass (10 seeds, cap = 5,000).

| Class | Mean recall | Std (pp) | 95% CI | Test samples |
|---|---|---|---|---|
| Normal | 94.1% | 2.27 | [92.5, 95.7] | 1,074,707 |
| Brute force | 100.0% | 0.02 | [99.9, 100.0] | 63,245 |
| Query flooding | 95.9% | 0.66 | [95.5, 96.4] | 597 |
| Frame stacking | 99.0% | 1.50 | [97.9, 100.0] | 134 |
| Replay | 7.9% | 4.46 | [4.7, 11.1] | 179 |
| Payload inj. | 94.6% | 0.00 | [94.6, 94.6] | 37 |
| Recon | 98.3% | 1.43 | [97.3, 99.4] | 36 |
| FDI | 100.0% | 0.00 | [100.0, 100.0] | 6 |
| Length manip. | 3.3% | 10.54 | n/a | 3 |

At cap = 5,000, Normal recall is 94.1%, meaning 5.9% of Normal traffic is incorrectly flagged. On 1,074,707 Normal test packets this corresponds to roughly 63,000 false alarms. In safety-critical OT environments, a missed attack carries the risk of equipment damage, process disruption or cascading failure across interconnected systems. A false alarm, by contrast, incurs an investigation cost that can be managed through alert aggregation and operator-facing filtering. This asymmetry is well established in OT security practice and makes recall-oriented detection a defensible design choice for a first-stage classifier whose outputs would feed into higher-level correlation and triage.

### 5.3 Cap Sensitivity and Approach Convergence

A cap sweep (500 to 50,000) validated across 10 seeds for Approaches 2b, 3 and 3b confirmed that the performance gap between approaches closes with increasing data volume (Fig. 5). At cap = 500, Approach 2b (79.4% ± 6.0pp) leads Approach 3 (64.0% ± 8.6pp) by 15.4 percentage points. By cap = 50,000, all three approaches converge: Approach 2b achieves 96.3% ± 1.6pp, Approach 3 achieves 96.6% ± 2.0pp and Approach 3b achieves 95.8% ± 2.8pp, with no pairwise difference reaching statistical significance. This convergence indicates that TCP/IP headers compensate for data scarcity at low caps but carry no independent discriminative signal once sufficient application-layer training data is available.

However, overall accuracy obscures per-class trade-offs. For Approach 2b, replay recall declines from 22.3% at cap = 500 to 2.8% at cap = 50,000 as the model increasingly classifies ambiguous packets as Normal. The optimal operating point therefore depends on whether deployment priorities favour overall accuracy or minority class coverage.

Approach 3 exhibits high variance across seeds at all cap values (standard deviation 6.7 to 9.6pp for caps below 10,000), confirming that the instability observed at cap = 5,000 is a systematic property of the 8-byte representation rather than an isolated training pathology. Only 2 of 20 runs at cap = 5,000 produced accuracy below 70%, both with seed 42, but the variance remains substantially higher than Approaches 2b or 3b across the full sweep.

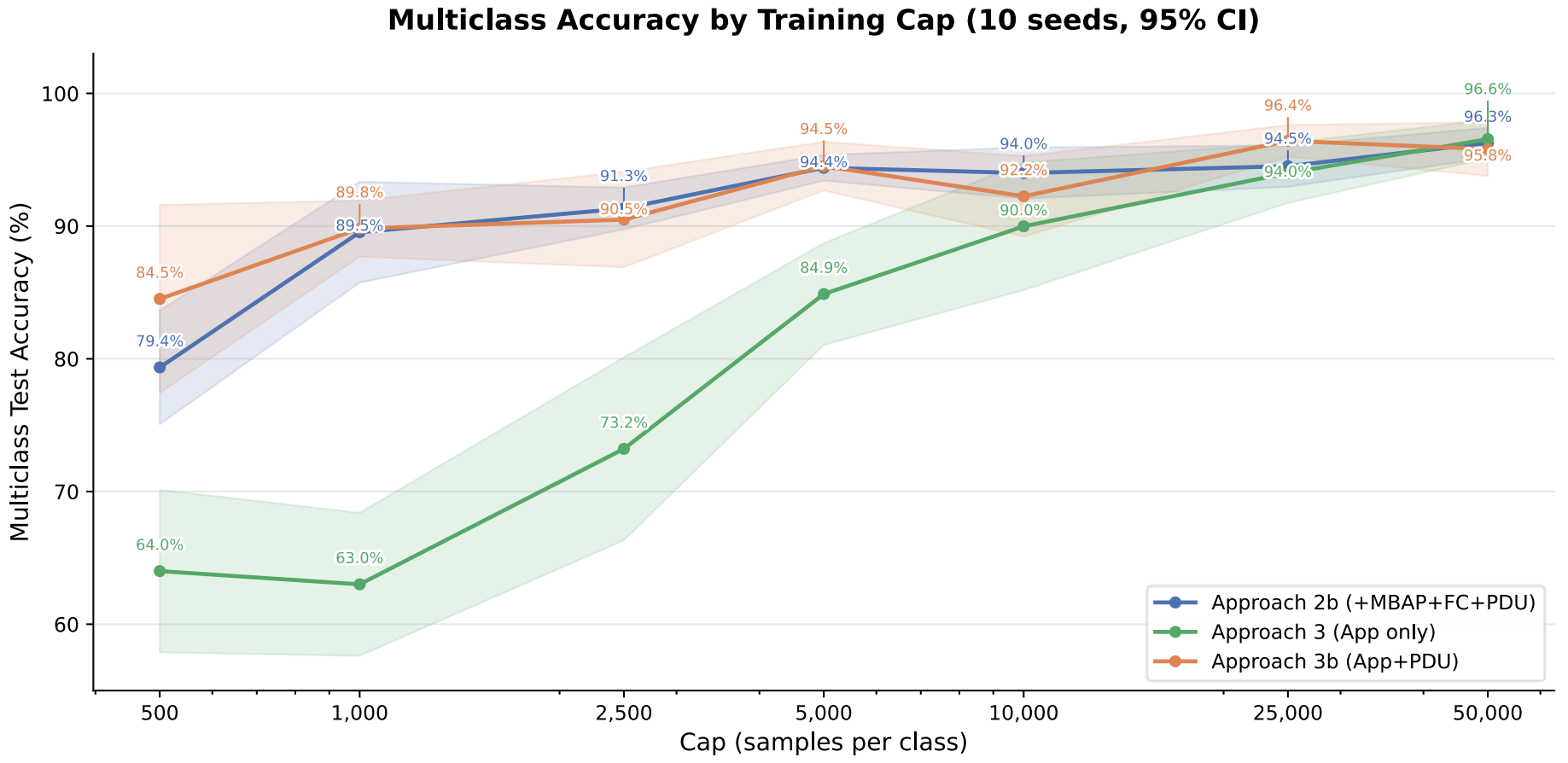

**Fig. 5.** Multiclass accuracy by training cap for Approaches 2b, 3 and 3b (10 seeds per point, shaded regions show 95% confidence intervals).

## 5.4 Hyperparameter Sensitivity

A $2^4$ factorial experiment (32 single-seed runs) confirmed that binary classification is insensitive to training configuration (0.02pp spread). For multiclass, the factorial identified Sigmoid activation with batch normalisation as the best single-seed configuration (95.9%). However, multi-seed validation (10 seeds) produced a mean of 93.2% ± 2.4pp. A paired t-test against the Phase 1 baseline showed no significant improvement ($p = 0.282$). This indicates that for these small CNNs, training configuration has a smaller effect than input representation. The five-approach comparison produces statistically significant differences between protocol depth levels, whereas hyperparameter optimisation does not reliably improve over the baseline. Convergence behaviour varied substantially across seeds: the epoch producing the best validation loss ranged from 6 to 79 (mean 32), indicating sensitivity to weight initialisation despite the low parameter count.

### 5.5 Comparison with Existing Work

Approach 2b achieves 98.6% binary accuracy on CIC Modbus 2023, a 14.4pp improvement over Russo, Zanasi and Marasco (2024) on the same dataset. Our 94.4% multiclass result uses 56,873 parameters, roughly 430 times fewer than the ResNet50-based approach of Kotsiopoulos et al. (2025). Direct accuracy comparison is not possible (different datasets, granularities and attack taxonomies), but the contribution is a demonstration that effective detection is achievable with a fraction of the computational cost. Compared with El-Sherif, Khattab and El-Soudani (2025), our Approach 1 result of 51.8% quantifies the challenge of transferring image-based methods from IoT to OT without adapting the input representation to include protocol-aware features. The following section examines why this gap exists and what it implies for OT deployment.

## 6 Discussion

### 6.1 Protocol Depth and the IoT-OT Gap

The most significant finding is the failure of TCP/IP-only classification on Modbus traffic (51.8%), contrasting with 100% on IoT. The explanation lies in OT traffic uniformity: near-identical headers for benign and malicious packets leave the CNN with no discriminative features at the TCP/IP layer. The Approach 1 multiclass result provides a diagnostic: frame stacking (97.8% recall) is detected because it modifies TCP-level packet length, confirming the CNN extracts whatever signal is present. This corroborates Anthi et al.'s (2021) finding that function code and frame length are the most important Modbus features. The single-packet classifier is designed as one component within a layered OT security architecture, with per-packet outputs feeding into higher-level temporal or process-aware monitoring.

For binary classification, eight bytes of application-layer information are both necessary and sufficient: accuracy jumps from 51.8% to 98.1% and gains negligibly thereafter (for stable approaches). For multiclass, the PDU operands add significant value: they enable FDI detection (100% recall for Approaches 2b and 3b, 0% for Approach 3). Approaches 2b and 3b are statistically indistinguishable at all cap values tested ($p > 0.05$ at six of seven caps), confirming that the TCP/IP headers provide no measurable advantage when PDU operands are included. At cap = 50,000, all three approaches converge to approximately 96% with no significant pairwise differences, reinforcing the interpretation that application-layer information is sufficient given adequate training data. The 8-byte Approach 3 representation remains the least stable, with the highest variance across seeds at every cap value.

### 6.2 Attack Detectability Limits

Three categories of detectability emerge. First, six attack types are reliably detected with recall above 94% at the recommended operating point: brute force, query flooding, frame stacking, payload injection, reconnaissance and FDI. Second, replay is structurally undetectable: replayed packets are copies of legitimate traffic with identical headers. The modest recall (7.9%) does not reflect genuine capability. Detecting replay would require temporal or sequence-based analysis. Third, length manipulation is undetectable due to insufficient samples (20 training, 3 test). Delay response was excluded entirely as a purely temporal attack. A comprehensive OT IDS would need to combine per-packet classification with complementary temporal methods; the design of such multi-stage architectures is beyond this paper's scope but is a natural direction for future work.

### 6.3 Parameter Efficiency

The binary classifiers evaluated in this work use 38 to 73 parameters; the multiclass classifier uses 56,873. By comparison, Kotsiopoulos et al. (2025) use over 23 million. This difference is consistent with deployment on embedded OT devices, though inference latency and throughput at realistic packet rates have not been measured on constrained hardware.

### 6.4 Limitations

The labelling methodology uses domain knowledge for ground truth only, but the rules are dataset-specific. All experiments use a single simulated dataset (CIC Modbus 2023) and a single protocol (Modbus TCP). All three attack scenarios were combined for training; cross-scenario generalisation has not been tested. The false positive rate (5.9% of Normal traffic flagged) would require alert aggregation in production. Alternative loss functions (focal loss) and regularisation (dropout) remain unevaluated.

The labelling pipeline has not been independently validated against a second annotator or an alternative methodology. While the rules are transparent and reproducible, the absence of inter-annotator agreement means the ground truth itself carries uncertainty. The recall-oriented evaluation strategy reflects the asymmetric cost structure of OT environments, where undetected attacks pose physical safety risks. This prioritisation of false negatives over false positives is consistent with established practice in safety-critical systems but may not be appropriate for all deployment contexts.

## 7 Conclusion

This paper extended the SPHBI methodology from IoT to OT, evaluating five approaches on 11.4 million Modbus TCP packets. TCP/IP headers alone are insufficient (51.8% binary), but eight bytes of application-layer information restore accuracy to 98.1%. The binary classifier (73 parameters for Approach 2b, 39 for Approach 3b) is stable across seeds (std ≤ 0.0pp) and represents a practical candidate for per-packet edge deployment. The best approach (2b) achieves 94.4% ± 2.2pp multiclass accuracy with 56,873 parameters, roughly 430 times fewer than ResNet50-based alternatives. Seven of eight detectable attack types are identified with recall above 94%. Replay and delay response are fundamental limits of the single-packet paradigm. Among protocol-aware approaches, Approaches 2b and 3b are statistically indistinguishable across all cap values, and all three approaches converge to approximately 96% accuracy at cap = 50,000. Approach 2b exhibits the lowest variance across seeds, making it the most reliable choice for deployment. Approach 3 (8 bytes, application layer only) is unstable across seeds in both tasks.

Future work will extend the methodology to DNP3, evaluate regularisation techniques, apply Grad-CAM attribution to identify which byte positions the CNN attends to for each attack type and validate inference on embedded OT hardware.

**Acknowledgements**. The author thanks colleagues at BMT. Marco Casassa Mont provided early direction for this research. Alec Wilson provided initial guidance, detailed feedback and constructive criticism on multiple drafts. Thomas Pett reviewed the paper and provided detailed feedback from an OT security perspective. Ryan Menzies reviewed the paper and provided feedback. Soma Maroju encouraged and provided access to the Databricks environment. Computational resources were provided by BMT.

**LLM usage.** Portions of this work were supported by Claude (Anthropic). The tool was used to assist with drafting and iterating manuscript text, generating figure code and structuring the multi-seed validation experiments. The experimental design, all methodological decisions and all scientific conclusions are the author's own. All experimental work was conducted independently by the author, who takes full responsibility for the content of this paper.